\documentclass[letterpaper]{JHEP3}

\newcommand{\beq}{\begin{equation}}
\newcommand{\eeq}{\end{equation}}
\newcommand{\cA}{{\cal A}}
\newcommand{\cAb}{{\overline{\cal A}}}
\newcommand{\cF}{{\cal F}}
\newcommand{\cFb}{{\overline{\cal F}}}
\newcommand{\cD}{{\cal D}}
\newcommand{\cDb}{{\overline{\cal D}}}
\newcommand{\cQ}{{\cal Q}}
\newcommand{\cU}{{\cal U}}
\newcommand{\cUb}{{\overline{\cal U}}} 
\newcommand{\KD}{{K\"{a}hler-Dirac }}
\newcommand{\Tr}{{\rm Tr\;}}
\newcommand{\etab}{{\overline{\eta}}}
\newcommand{\psib}{{\overline{\psi}}}
\newcommand{\phib}{{\overline{\phi}}}
\newcommand{\bx}{{\bf x}}
\newcommand{\bmu}{{\boldsymbol \mu}}
\newcommand{\bnu}{{\boldsymbol \nu}}

\newcommand{\be}{{\boldsymbol e}}
\newcommand{\bzero}{{\boldsymbol 0}}
\usepackage{graphicx}
\usepackage{amsmath}
\usepackage{amsfonts}
\usepackage{epsfig}

\title{From Twisted Supersymmetry to Orbifold Lattices}

\author{Simon Catterall \\
Department of Physics, Syracuse University, Syracuse, NY13244, USA \\
E-mail: \email{smc@phy.syr.edu}
}

\preprint{}

\date{November 2007}

\abstract{
We show how to derive the supersymmetric orbifold lattices of
Cohen et al. \cite{Cohen:2003xe,Cohen:2003qw} and Kaplan 
et al. \cite{Kaplan:2005ta} by direct 
discretization
of an appropriate twisted supersymmetric Yang-Mills theory. We examine in
detail the four supercharge two dimensional theory and the theory with
sixteen supercharges in
four dimensions. The continuum limit of the latter theory is the well
known Marcus twist of ${\cal N}=4$ Yang-Mills. The lattice models are
gauge invariant and possess one
exact supersymmetry at non-zero lattice spacing. 
}

\begin{document}

%-----------------------------------------------------------------
%
\section{Introduction}
%
%-----------------------------------------------------------------
The problem of putting supersymmetry on the lattice 
is an old one going back at least 25 years 
(see the review \cite{Feo_rev2} and references therein).
However most of the older work utilized discretization schemes
that break supersymmetry completely at the classical level.
With a few notable exceptions like ${\cal N}=1$ super Yang-Mills
in $D=4$ such an approach generically leads 
to fine tuning problems -- the
couplings to a set of induced SUSY violating operators must be
tuned carefully to zero as the lattice spacing is reduced \cite{Curci:1986sm}. 
In low dimensions this fine tuning may be manageable 
since the theories are
super-renormalizable and hence all divergences occur in low
orders of perturbation theory \cite{Golterman_wz,Elliott:2005bd}. 

Recently, however, the field has seen a resurgence of activity
due to the
realization that a certain subclass of theories could be
discretized while preserving a fraction of the continuum
supersymmetries \cite{Kaplan:2002wv,
Cohen:2003xe,Cohen:2003qw,Kaplan:2005ta,Catterall_topo,Catterall_wz1,
Catterall_sig1}. Two main approaches have been followed; obtaining
a lattice theory by orbifolding a supersymmetric matrix model
(for a good review of this approach see \cite{Giedt_rev1}) and
direct discretization of a reformulation of theory in terms of twisted
fields\footnote{A third approach based on deformation of the IIB matrix model
appears to provide an independent construction of the ${\cal N}=4$ theory which
also preserves a scalar supersymmetry
\cite{Unsal:2005us}}. The twisting procedure goes back to Witten \cite{Witten:1988ze}
in his seminal construction of topological field theories
but actually had been anticipated in earlier lattice work using
\KD fields \cite{Elitzur:1982vh}. The precise connection between
the \KD fermion mechanism and topological twisting was found by
Kawamoto and collaborators \cite{Kawamoto:1999zn,Kato_bf}. 

While the orbifold constructions are essentially unique
\cite{Damgaard_orb} various
approaches to discretization of the twisted theories have been
advocated in \cite{Catterall_n=2, Sugino_sym1,Sugino_2d,Sugino_sym2,
D'Adda_super,D'Adda_2d}. Recent work by Damgaard, Matsuura and
Takimi has indicated that there are, in fact, strong connections
between these twisted theories and the orbifold models
\cite{Damgaard:2007xi,Damgaard:2007eh, Takimi:2007nn}. This had already
been anticipated by Unsal who showed that the naive continuum
limit of the sixteen supercharge orbifold model
in four dimensions led to the Marcus twist of ${\cal N}=4$
Yang-Mills \cite{Unsal_rel}.

In this paper we complete this web of interconnections by showing that
the orbifold actions can be obtained by direct discretization of an appropriate
twist of the supersymmetric Yang-Mills theory. We consider first
the two dimensional theory with four supercharges which has been extensively
discussed in the literature and for which numerical simulations have
already been attempted \cite{Catterall_sims, Catterall_rest, Kanamori:2007yx}.
We then show how to rewrite the continuum Marcus twist of ${\cal N}=4$
as the dimensional
reduction of a very simple five dimensional theory which is almost
of the same form as the two dimensional theory. 
This simple five dimensional structure allows us 
to use the geometric discretization prescription
employed in two dimensions to write down a supersymmetric
lattice theory corresponding
to this Marcus twist of ${\cal N}=4$ Yang-Mills. The resulting theory is nothing more
than the $\cQ=16$ orbifold lattice theory in four dimensions.

\section{Four supercharge theory in two dimensions}
\subsection{Continuum twisted theory}
 
Following the arguments given in \cite{Catterall:2005eh, Kato_bf,
Kawamoto:1999zn} the 
continuum theory
is first rewritten in {\it twisted} form. 
The bosonic
sector of the twisted theory comprises a single {\it complexified}
gauge connection $\cA$ and a scalar auxiliary field $d$. Fermionic degrees of
freedom are naturally embedded as components of a single complex
\KD field $\Psi=(\eta,\psi_\mu,\chi_{\mu\nu})$
whose components are antisymmetric tensor fields. We will
take all these fields as living in the adjoint of a $U(N)$ gauge group.
The twisted theory naturally
possesses a nilpotent scalar supercharge $\cQ$ whose action on these
fields is given by 
\begin{eqnarray}
\cQ\; \cA_\mu&=&\psi_\mu\nonumber\\
\cQ\; \psi_\mu&=&0\nonumber\\
\cQ\; \cAb_\mu&=&0\nonumber\\
\cQ\; \chi_{\mu\nu}&=&-\cFb_{\mu\nu}\nonumber\\
\cQ\; \eta&=&d\nonumber\\
\cQ\; d&=&0
\end{eqnarray}
Notice that in this formulation of the twisted theory all the physical
bosonic
degrees of freedom are carried by the complex gauge field. The scalar
supercharge that is employed here corresponds to a complex combination of
the scalar supercharge $Q$ employed in earlier twisted lattice constructions
\cite{Catterall_n=2, Catterall_n=4,Sugino_sym1, Sugino_2d}  and its 2-form dual
$Q_{12}$. Notice that this supersymmetry implies that the fermions are complex
which is natural in a Euclidean theory.
As in previous constructions the twisted action in two dimensions can be written
in $\cQ$-exact form $S=\beta\cQ\; \Lambda$ where $\Lambda$ is
\beq
\Lambda=\int
\Tr\left(\chi_{\mu\nu}\cF_{\mu\nu}+\eta [ \cDb_\mu,\cD_\mu ]-\frac{1}{2}\eta
d\right)
\eeq
and we have introduced the complexified covariant derivatives (we employ
an antihermitian basis for the generators of $U(N)$)
\begin{eqnarray}
\cD_\mu&=&\partial_\mu+\cA_\mu=\partial_\mu+A_\mu+iB_\mu\nonumber\\
\cDb_\mu&=&\partial_\mu+\cAb_\mu=\partial_\mu+A_\mu-iB_\mu
\end{eqnarray}
Doing the $\cQ$-variation and integrating out the field $d$ yields
\beq
S=\int\Tr \left(-\cFb_{\mu\nu}\cF_{\mu\nu}+\frac{1}{2}[ \cDb_\mu, \cD_\mu]^2-
\chi_{\mu\nu}\cD_{\left[\mu\right.}\psi_{\left.\nu\right]}-\eta \cDb_\mu\psi_\mu\right)\eeq
The bosonic terms can be written
\begin{eqnarray}
\cFb_{\mu\nu}\cF_{\mu\nu}&=&\left(F_{\mu\nu}-[B_\mu,B_\nu]\right)^2+
\left(D_{\left[\mu\right.}B_{\left.\nu\right]}\right)^2\nonumber\\
\frac{1}{2}\left[\cDb_\mu,\cD_\mu\right]^2 &=& -2\left(D_\mu B_\mu\right)^2
\end{eqnarray}
where $F_{\mu\nu}$ and $D_\mu$ denote the usual field strength and
covariant derivative depending on the real part of the connection $A_\mu$.
After integrating by parts the term linear in $F_{\mu\nu}$ cancels
and the final bosonic action reads\footnote{The bosonic action is real positive
definite on account of the antihermitian basis that we have chosen}
\beq
S_B=\int\Tr \left(-F^2_{\mu\nu}+2B_\mu D_\nu D_\nu B_\mu-[B_\mu,B_\nu]^2\right)\eeq
Notice that the imaginary parts of the gauge field have transformed into
the two scalars of the super Yang-Mills theory! This is further confirmed
by looking at the fermionic part of the action which can be rewritten
in $2\times 2$ block form as
\beq
\left(\begin{array}{cc}\chi_{12}&\frac{\eta}{2}\end{array}\right)
\left(\begin{array}{cc}-D_2-iB_2&D_1+iB_1\\
                        D_1-iB_1&D_2-iB_2\end{array}\right)
\left(\begin{array}{c}\psi_1\\ \psi_2\end{array}\right)
\eeq
which is easily recognized as the dimensional reduction of ${\cal N}=1$
super Yang-Mills theory in four dimensions in which a chiral representation
is employed for the fermions. As usual the scalar fields $B_\mu$ arise from
the gauge fields in the reduced directions.

\subsection{Lattice theory}

The transition to the lattice theory is straightforward; we employ the
geometrical discretization scheme proposed in \cite{Catterall_n=2}. For
completeness we summarize it here.
In general continuum p-form fields are mapped
to lattice fields defined on p-subsimplices of a general
simplicial lattice. In the case of hypercubic lattices
this assignment is equivalent to placing a p-form with
indices $\mu_1\ldots\mu_p$ on the link connecting $\bx$ with
$(\bx+\bmu_1+\ldots+\bmu_p)$ where $\bmu_i,i=1\ldots p$ corresponds to
a unit vector in the lattice. 
Actually this is not quite the full story; each link has two
possible orientations and we must also specify which orientation
is to be used for a given field. A positively oriented field
corresponds to one in which the link vector has positive components with
respect to the coordinate basis.

Continuum derivatives on such a hypercubic
lattice are represented by lattice difference operators acting on these
link fields. Specifically,
covariant derivatives appearing in curl-like operations 
and {\it acting on positively oriented fields} are replaced by
a lattice gauge covariant forward difference operator whose action on
lattice scalar and vector fields is given by
\begin{eqnarray}
\cD^{(+)}_\mu f(\bx)=\cU_\mu(\bx)f(\bx+\bmu)-f(\bx)\cU_\mu(\bx)\nonumber\\
\cD^{(+)}_\mu f_\nu(\bx)=\cU_\mu(\bx)f_\nu(\bx+\bmu)-f_\nu(\bx)\cU_\mu(\bx+\bnu)
\label{derivs}
\end{eqnarray}
where $\bx$ denotes a two dimensional lattice vector and
$\bmu=(1,0)$, $\bnu=(0,1)$ unit
vectors in the two coordinate directions.
Here, we have replaced the continuum {\it complex}
gauge fields $\cA_\mu$ by non-unitary link fields $\cU_\mu=e^{i\cA_\mu}$. 
The backward difference
operator $\cDb^-_\mu$ replaces the continuum covariant derivative
in divergence-like operations and its action on (positively oriented)
lattice vector fields can be
gotten by requiring that it be adjoint to $\cD^+_\mu$. Specifically
its action on lattice vectors is
\beq
\cDb^{(-)}_\mu f_\mu(\bx)=f_\mu(\bx)\cUb_\mu(\bx)-
\cUb_\mu(\bx-\bmu)f_\mu(\bx-\bmu)
\eeq
The nilpotent scalar supersymmetry now acts on the lattice fields
as
\begin{eqnarray}
\cQ\; \cU_\mu&=&\psi_\mu\nonumber\\
\cQ\; \psi_\mu&=&0\nonumber\\
\cQ\; \cUb_\mu&=&0\nonumber\\
\cQ\; \chi_{\mu\nu}&=&\cF^{L\dagger}_{\mu\nu}\nonumber\\
\cQ\; \eta&=&d\nonumber\\
\cQ\; d&=&0
\end{eqnarray}
Here we written the lattice field strength as 
\beq
\cF^L_{\mu\nu}=\cD^{(+)}_\mu
\cU_\nu(\bx)=\cU_\mu(\bx)\cU_\nu(\bx+\bmu)-
\cU_\nu(\bx)\cU_\mu(\bx+\bnu)\label{field}\eeq
which
reduces to the continuum (complex) field strength in the naive continuum
limit and is automatically antisymmetric in the indices $(\mu,\nu)$.

Notice that this supersymmetry transformation
implies that the fermion fields $\psi_\mu$ have 
the same orientation as their superpartners
the gauge links $\cU_\mu$ and run from $\bx$ to $(\bx+\bmu)$. However, the
field $\chi_{\mu\nu}$ must have
the same orientation as $\cF^{L\dagger}_{\mu\nu}$ and
hence is to be assigned to the 
negatively oriented link running from $(x+\bmu+\bnu)$ down to $\bx$ 
i.e parallel to the vector $(-1,-1)$. This link choice also follows naturally from
the matrix representation of the \KD field $\Psi$
\beq
\Psi=\eta I+\psi_\mu \gamma_\mu +\chi_{12}\gamma_1\gamma_2\eeq
which associates the field $\chi_{12}$ with the
lattice vector $\bmu_1+\bmu_2=\bmu+\bnu$. We will see that
the negative orientation is crucial for allowing us
to write down gauge invariant expressions for the fermion kinetic
term.
Finally, it should be clear that the scalar fields
$\eta$ and $d$ can be taken to transform simply as site fields.

These link mappings and orientations are conveniently summarised by
giving the gauge transformation properties of the lattice fields
\begin{eqnarray}
\eta(\bx)&\to&G(\bx)\eta(\bx)G^\dagger(\bx)\nonumber\\
\psi_\mu(\bx)&\to&G(\bx)\psi_\mu(\bx)G^\dagger(\bx+\bmu)\nonumber\\
\chi_{\mu\nu}(\bx)&\to&G(\bx+\bmu+\bnu)\chi_{\mu\nu}G^\dagger(\bx)\nonumber\\
\cU_\mu(\bx)&\to&G(\bx)\eta(\bx)G^\dagger(\bx)\nonumber\\
\cUb_\mu(\bx)&\to&G(\bx+\bmu)\cUb_\mu(\bx)G^\dagger(\bx)
\end{eqnarray}
Notice that this choice of link and
orientation for the twisted lattice fields maps exactly
into their r-charge assignments in the orbifolding approach
\cite{Cohen:2003xe}. Furthermore,
the above $\cQ$-variations and field assignments are equivalent to the
approach described in \cite{D'Adda_2d} {\it provided} that we set the fermionic
shift parameter $a$ in that formulation to zero and consider only
the corresponding scalar superymmetry. 

The lattice gauge fermion now takes the form
\beq
\Lambda=\sum_{\bx}\Tr\left( \chi_{\mu\nu}\cD^{(+)}_\mu\cU_\nu+\eta 
\cDb^{(-)}_\mu \cU_\mu-\frac{1}{2}\eta d\right)\eeq
It is easy
to see that in the naive continuum limit the lattice
divergence $\cDb^{(-)}_\mu \cU_\mu$ equals 
$[\cDb_\mu,\cD_\mu]$. Notice that with the
previous choice of orientation for the various fermionic link fields
this gauge fermion is automatically invariant under lattice gauge
transformations. There is no need for the doubling of degrees of freedom
necessary in previous approaches to geometric discretization
\cite{Catterall_n=2,Catterall_n=4}. In those constructions the
nature of the gauge fermion and the scalar supercharge led to the
presence of explicit Yukawa interactions in the theory. These
in turn required the lattice theory to contain fermion link fields
of both orientations and hence led to a doubling of degrees of
freedom with respect to the continuum theory. In the
twist described in this paper the Yukawa interactions are embedded
into the complexified covariant derivatives and
successive
components of the \KD field representing the fermions
can be chosen with alternating orientations leading to a \KD action
which is automatically gauge invariant without these extra degrees of
freedom.

Acting with the $\cQ$-transformation shown above and again integrating out
the auxiliary field $d$ we derive the gauge and $\cQ$-invariant lattice
action
\beq
S=\sum_{\bx}\Tr\left(\cF^{L\dagger}_{\mu\nu}\cF^L_{\mu\nu}+
\frac{1}{2}\left(\cDb^{(-)}_\mu \cU_\mu\right)^2-
\chi_{\mu\nu}\cD^{(+)}_{\left[\mu\right.}\psi_{\left.\nu\right]}-
\eta \cDb^{(-)}_\mu\psi_\mu\right)
\eeq
But this is precisely the orbifold action arising in \cite{Cohen:2003xe}
with the modified deconstruction step described in \cite{Unsal_comp} and
\cite{Damgaard:2007xi}. 
The two approaches are thus entirely equivalent. 

We can use this geometrical formulation to show very easily 
that the lattice theory exhibits no fermion doubling problems. The simplest
way to do this 
is merely to notice that the lattice action at zero coupling $\cU\to I$
conforms to the
canonical form required for no doubling by the theorem of 
Rabin \cite {Rabin:1981qj}.
Explicitly, discretization of continuum geometrical actions
will not encounter doubling problems if
continuum derivatives acting in curl-like operations are replaced by forward
differences in the lattice theory while
continuum derivatives appearing in divergence-like operations are represented
by backward differences on the lattice. More precisely the continuum
exterior derivative $d$ is mapped to a forward difference while its adjoint
$d^\dagger$ is represented by a backward difference.

However we can also see this by simply examining the 
the form of the fermion operator arising in this construction.
\beq
\left(\begin{array}{cc}\chi_{12}&\frac{\eta}{2}\end{array}\right)
\left(\begin{array}{cc}-\cD^{(+)}_2&\cD^{(+)}_1\\
                        \cD^{(-)}_1&\cD^{(-)}\end{array}\right)
\left(\begin{array}{c}\psi_1\\ \psi_2\end{array}\right)
\eeq
Clearly the determinant of this operator in the free limit is nothing more
than the usual determinant encountered for scalars in two dimensions and
hence possesses no extraneous zeroes that survive the continuum limit.

\section{Sixteen supercharge theory in four dimensions}
\subsection{Continuum twisted theory}

In the $\cQ=4$ theory in two
dimensions the physical degrees of freedom were encoded in a complex
gauge field. The same idea applied to the $\cQ=16$ theory naturally
leads us to consider a theory of complex gauge fields
$\cA_a, a=1\ldots 5$ in {\it five
dimensions}. Paralleling the four supercharge theory we introduce
an additional
auxiliary bosonic scalar field $d$ and a set of five dimensional
antisymmetric tensor fields 
to represent the fermions
$\Psi=(\eta,\psi_a,\chi_{ab})$. This latter field
content corresponds to considering just one of the two
\KD fields used to represent the 32 fields of the five dimensional
theory. Again, a nilpotent symmetry relates
these fields
\begin{eqnarray}
\cQ\; \cA_a&=&\psi_a\nonumber\\
\cQ\; \psi_a&=&0\nonumber\\
\cQ\; \cAb_a&=&0\nonumber\\
\cQ\; \chi_{ab}&=&-\cFb_{ab}\nonumber\\
\cQ\; \eta&=&d\nonumber\\
\cQ\; d&=&0
\end{eqnarray}
and we may write down the same $\cQ$-exact action that was employed in
two dimensions $S=\beta\cQ \Lambda$ with
\beq
\Lambda=\int
\Tr\left(\chi_{ab}\cF_{ab}+\eta [ \cDb_a,\cD_a ]-\frac{1}{2}\eta
d\right)
\eeq
where we have again employed complexified covariant derivatives.
Carrying out the $\cQ$-variation and subsequently integrating out the
auxiliary field as for the $\cQ=4$ theory leads to the action
\beq
S=\int\Tr \left(-\cFb_{ab}\cF_{ab}+\frac{1}{2}[ \cDb_a, \cD_a]^2-
\chi_{ab}\cD_{\left[a\right.}\psi_{\left.b\right]}-\eta \cDb_a\psi_a\right)\eeq
Actually in this theory there is another fermionic term one can write down which is also
invariant under this supersymmetry taking the form
\beq
S_{\rm closed}=-\frac{1}{2}\int\epsilon_{abcde}\chi_{ab}\cDb_c\chi_{de}\eeq
The invariance of this term is just a result of the Bianchi
identity $\epsilon_{abcde}\cD_c \cF_{de}=0$. The final action we will
employ is the sum of the $\cQ$-exact piece and this $\cQ$-closed term.
The coefficient in front of this term is determined by the requirement
that the theory reproduce the Marcus twist of ${\cal N}=4$ Yang-Mills.

Clearly to make contact with a twist of ${\cal N}=4$ in four dimensions
we must dimensionally reduce this theory along the 5th direction.
This will yield a complex scalar $\phi=A_5+iB_5$ and its superpartner $\etab$.
The 10 five dimensional fermions $\chi_{ab}$ 
naturally decompose into a 2-form
$\chi_{\mu\nu}$ and vector $\psib_{\mu}$ in four dimensions. 
\begin{eqnarray}
\cA_a&\to&\cA_\mu\oplus\phi\nonumber\\
\cF_{ab}&\to& \cF_{\mu\nu}\oplus\cD_\mu \phi\nonumber\\
\left[\cDb_a,\cD_a\right]&\to&\left[\cDb_\mu,\cD_\mu\right]\oplus
\left[\phib,\phi\right]\nonumber\\
\psi_a&\to&\psi_\mu\oplus\etab\nonumber\\
\chi_{ab}&\to&\chi_{\mu\nu}\oplus\psib_{\mu}
\end{eqnarray}
where we will employ the convention that Greek indices run from one to
four and are reserved for
four dimensional tensors while Roman indices refer to the original
five dimensional theory. The reduced action takes the form
\begin{eqnarray}
S&=&\int\Tr \left(-\cFb_{\mu\nu}\cF_{\mu\nu}+\frac{1}{2}\left[\cDb_\mu,
\cD_\mu\right]^2+
\frac{1}{2}\left[\phib ,\phi\right]^2+(\cD_\mu\phi)^\dagger(\cD_\mu\phi)-
\chi_{\mu\nu}\cD_{\left[\mu\right.}\psi_{\left.\nu\right]}\right.\nonumber\\
&-&\left.\psib_\mu \cD_\mu\etab-\psib_\mu\left[\phi,\psi_\mu\right]
-\eta \cDb_\mu\psi_\mu -\eta\left[\phib,\etab\right]-
\chi^*_{\mu\nu}\cDb_\mu\psib_\nu-
\chi^*_{\mu\nu}\left[\phib,\chi_{\mu\nu}\right]\right)
\end{eqnarray}
where the last two terms arise from dimensional reduction of the $\cQ$-closed
term and $\chi^*$ is the Hodge dual of $\chi$, 
$\chi_{\mu\nu}=\frac{1}{2}\epsilon_{\mu\nu\rho\lambda}\chi_{\rho\lambda}$.
Up to trivial rescalings this is the action (with gauge
parameter $\alpha=1$) of twisted ${\cal N}=4$ Yang-Mills in four
dimensions written down by Marcus \cite{Marcus}\footnote{It is also
the twist of ${\cal N}=4$ YM used in the Geometric Langlands program
\cite{Kapustin:2006pk}}. 
This twisted action is well known to be fully equivalent to the usual
form of ${\cal N}=4$ in flat space. Here, we have shown
how to derive this theory by dimensional reduction of a rather simple
five dimensional theory employing a complex gauge field and integer spin
twisted fermions. It will be the basis of our lattice formulation to which
we now turn.

\subsection{Lattice theory}

The discretization scheme we use is precisely the same as 
for two supercharge theory. Complex five dimensional gauge fields are replaced
by complex gauge links $\cU_a,a=1\ldots 5$. The $\cQ$-supersymmetry is
essentially the same as in the continuum and remains nilpotent
\begin{eqnarray}
\cQ\; \cU_a&=&\psi_a\nonumber\\
\cQ\; \cUb_a&=&0\nonumber\\
\cQ\; \psi_a&=&0\nonumber\\
\cQ\; \chi_{ab}&=&\left(\cF^L_{ab}\right)^\dagger\nonumber\\
\cQ\; \eta&=&d\nonumber\\
\cQ\; d&=&0
\label{latticeQ}
\end{eqnarray}
where the lattice field strength $\cF^L_{ab}$ is given by eqn.~\ref{field}
as before. The chief difficulty remaining is to decide how the
continuum tensor fields are to be assigned to lattice links after
dimensional reduction to four dimensions. For the moment let us base
our 
discretization scheme
around a hypercubic lattice.  Then the gauge links
$\cU_\mu\equiv \cU_a, a=1\ldots 4$ 
should live on elementary coordinate directions in the unit hypercube.
This then implies that the
superpartners of those gauge links $\psi_\mu$ should also live on
those links and be oriented 
in the same fashion i.e running from
$\bx$ to $(\bx+\bmu)$. We will adopt the notation that these four
basis vectors are labeled $\bmu_a, a=1\ldots 4$.

However the assignment of $\psi_5$  is not
immediately obvious -- a naive assignment to a site field
would result in {\it two} fermionic scalars which
is not what is expected for a four dimensional \KD field. The same
line of reasoning suggests in fact that we
associate $\psi_5$ with the 4-form component of that field.
An independent line of argument confirms this;
the field $\psi_5$ is part of the vector component of 
a five dimensional \KD field and is thus
associated with the five dimensional gamma matrix
$\Gamma^5$. This is usually represented by the chiral matrix of the
four dimensional theory $\Gamma^5=\gamma_5=\gamma_1\gamma_2\gamma_3\gamma_4$
and suggests a 4-form interpretation for the field in the four
dimensional theory. 
As for two dimensions this motivates assigning the lattice field to
the body diagonal of the unit hypercube. 
Actually we must be careful; this same
assignment will also apply to the field $\cU_5$. To construct
the bosonic action we need to able to apply $\cD^{(+)}_a$
to this link field and stay within the unit hypercube. 
Thus we choose the fields to be
oriented in the {\it opposite direction} corresponding to the
basis vector $\bmu_5=(-1,-1,-1,-1)$. Notice that
this assignment ensures that $\sum_{a=1}^5\bmu_a=\bzero$ which will be
seen to be crucial for constructing gauge invariant quantities.

As for two dimensions we will summarize
these link and orientation assignments
by writing down a set of gauge transformations for the fields 
\begin{eqnarray}
\eta(\bx)&\to& G(\bx)\eta(\bx) G^\dagger(\bx)\nonumber\\
\psi_a(\bx)&\to& G(\bx)\psi_a(\bx) G(\bx+\bmu_a)\nonumber\\
\chi_{ab}(\bx)&\to&G(\bx+\bmu_a+\bmu_b)\chi_{ab}(\bx)G^\dagger(\bx)\nonumber\\
\cU_a(\bx)&\to&G(\bx)\cU_\mu(\bx)G^\dagger(\bx+\bmu_a)\nonumber\\
\cUb_a(x)&\to&G(\bx+\bmu_a)\cUb_\mu(\bx)G^\dagger(\bx)\\
\end{eqnarray}
In this form the reader will see that the gauge transformations of fields
in this four dimensional theory follow almost exactly
the same form as their cousins in two dimensions. 
Notice also that these link choices and orientations
match exactly the r-charge assignments of the orbifold action for the sixteen
supercharge theory in four dimensions \cite{Kaplan:2005ta}. 

However, one should note that with these conventions not all fields lie in the
positively oriented unit hypercube. The problematic fields all possess
a tensor index $a=5$. However they can be mapped into the hypercube by
a simple lattice translation. The transformation is
\begin{eqnarray}
\chi_{5\mu}(\bx-\bmu_5-\bmu)&\to&\frac{1}{3!}\epsilon_{\mu\nu\lambda\rho}
\theta_{\nu\lambda\rho}(\bx)\nonumber\\
\psi_5(\bx-\bmu_5)&\to&\frac{1}{4!}\epsilon_{\mu\nu\lambda\rho}
\kappa_{\mu\nu\lambda\rho}(\bx)
\label{change}
\end{eqnarray}
where we have relabeled the mapped fields so as to match their corresponding
link assignment in the unit hypercube. Notice that $\chi_{5\mu}$ contains the
field $\theta_{\nu\lambda\rho}$ which plays the role of the 3-form component
of a four dimensional \KD field. The 2-form and 4-form components are then 
supplied by $\chi_{ab},a,b=1\ldots 4$ and
$\kappa_{\mu\nu\lambda\rho}$. Furthermore
the $\theta_{\nu\lambda\rho}$ and $\kappa_{\mu\nu\lambda\rho}$ fields
have positive and negative orientation.
Thus, as for two dimensions, successive components of the
resultant fermionic \KD field alternate in orientation which will be the
key to writing down gauge invariant fermion kinetic terms. 
Clearly any expression which is summed over all lattice points will
be invariant under such a translation and we will use this freedom
later to recast the lattice action in a way which makes clear why the
fermionic action does not suffer from doubling problems.

The advantage of the 5D variables is that they allow easy comparison
with the analogous orbifold expressions and are compatible with our
previous expressions for gauge covariant finite differences which can now
be written in the general form
\begin{eqnarray}
\cD^{(+)}_c f_d(\bx)&=&\cU_c(\bx)f_d(\bx+\bmu_c)-f_d(\bx)U_c(\bx+\bmu_d)\\
\cD^{(-)}_c f_c(\bx)&=&f_c(\bx)\cUb_c(\bx)-\cUb_c(x-\bmu_c)f_c(\bx-\bmu_c)
\end{eqnarray}
Using these ingredients the lattice action arising from the 
$\cQ$-exact piece of the
continuum action takes the form
\beq
S=\sum_{\bx}\Tr \left(\cF^{L\dagger}_{ab}\cF^{L}_{ab}+
\frac{1}{2}\left(\cDb^{(-)}_a \cU_a\right)^2-
\chi_{ab}\cD^{(+)}_{\left[a\right.}\psi_{\left.b\right]}-
\eta \cDb^{(-)}_a\psi_a\right)\eeq
There is one remaining subtlety in this identification. Exactly how
does the $\cQ$-closed term remain supersymmetric under discretization ?
A natural lattice analog of $\cDb_c\chi_{ab}$ 
is given by
\beq
\cDb^{(-)}_c\chi_{ab}(\bx)=\chi_{ab}(\bx)\cUb_c(\bx-\bmu_c)-
\cUb_c(\bx+\bmu_a+\bmu_b-\bmu_c)\chi_{ab}(\bx-\bmu_c)
\eeq
Using this it is straightforward to write down a gauge invariant
lattice analog of
the continuum $\cQ$-closed term 
\beq
S_{\rm closed}=-\frac{1}{2}\sum_{\bx}\Tr 
\epsilon_{abcde}\chi_{de}(\bx+\bmu_a+\bmu_b+\bmu_c)
\cD^{(-)}_c\chi(\bx+\bmu_c)\label{closed}\eeq
Notice that the $\epsilon$-tensor forces all indices to be distinct
and the gauge invariance of this result follows from
the fact that $\sum_{i=1}^5\bmu_i=\bzero$.
It is easy to see that it is equal to
third fermionic term of the
orbifold action appearing in eqn.~(3.18) of reference \cite{Kaplan:2005ta}. 

In the continuum the invariance of this term under $\cQ$-transformations
requires use of the
Bianchi identity.
Remarkably, the lattice difference operator
satisfies a similar identity (see \cite{Aratyn} for the four dimensional
result)
\beq
\epsilon_{abcde}D^{(+)}_c\cF^{L}_{de}=0\eeq
Thus the discretization of the $\cQ$-closed term in eqn.~\ref{closed}
is indeed
invariant under the lattice $\cQ$-transformation given in eqn.~\ref{latticeQ}.
This completes
the proof of the equivalence. The connection between the naive continuum
limit of the orbifold lattice 
and the Marcus twist of ${\cal N}=4$ super
Yang-Mills was shown earlier by Unsal \cite{Unsal_rel}; in this paper
we make this connection explicit by discretizing the latter theory in a way
which maintains the scalar supersymmetry and obtain the orbifold
action directly.

Finally to obtain the hypercubic lattice discretization 
of the continuum Marcus
theory requires setting $\cU_5=\phi$ a complex field
with vanishing expectation value. Notice though that
this discretization contains elementary links of varying
length. Actually the lattice action we have derived is 
clearly supersymmetric
for arbitrary deformations of the lengths
and orientations of the five basic vectors $\bmu_a,a=1\ldots 5$.
Thus it is possible to consider the symmetric situation in which the
lattice {\it in spacetime}
is constructed from a unit cell in which these basis
vectors are
equivalent -- they point out from the center
of a four-dimensional hypertetrahedron to its
five vertices. These vectors $e_i,i=1\ldots 5$
are given explicitly in \cite{Kaplan:2005ta}.
At the same time we must set $\cU_5$ to the exponential
of a complex matrix to maintain symmetry with the
other link fields $\cU_\mu,\mu=1\ldots 4$. 
This construction necessitates
introducing a map between the abstract lattice used to build the
supersymmetric theory and spanned by the
integer component vectors $\bx=(n_1,n_2,n_3,n_4)$ and the 
physical spacetime coordinates $\bf R$.
Explicitly ${\bf R}=\sum_{i=1}^5 n_i \be_i$. 
Such a lattice has the
point group symmetry $S^5$ which is much larger than the $S^4$ symmetry
of the hypercubic lattice - a factor which may prove to be important
when examining the restoration of rotational invariance and the
other supersymmetries of the continuum ${\cal N}=4$ theory.

\subsection{Absence of fermion doubling}

Finally this geometric approach makes it easier to understand why this
lattice theory does not suffer from doubling problems.
We will analyze this question in the context of the
hypercubic lattice discretization. Clearly most of
the fermionic kinetic terms manifestly satisfy the double free discretization 
prescription given by Rabin \cite{Rabin:1981qj}.
The only difficult terms arise
when one or more tensor indices of the fields equal $a=5$. Expressions involving
these fields are not located wholly in the positively oriented unit
hypercube and must be translated into the hypercube before they
can examined from the perspective of this prescription.
As an example,
consider the term
\beq
\sum_{\bx}\Tr\chi_{5\mu}D^{(+)}_\mu\psi_5=
\sum_{\bx}\Tr\chi_{5\mu}(\bx)\left(\cU_\mu(\bx)\psi_5(\bx+\bmu)-
\psi_5(\bx)\cU_\mu(\bx+\bmu_5)\right)
\eeq
We 
first shift the coordinates $\bx\to (\bx-\bmu_5-\bmu)$ and then 
use the previous change of variables given in eqn.~\ref{change}
to rewrite this as
\beq
\frac{1}{3!}\sum_{\bx}\Tr\theta_{\nu\lambda\rho}(\bx)\left(\cU_\mu(\bx-\bmu_5-\bmu)
\kappa_{\mu\nu\lambda\rho}(\bx)-
\kappa_{\mu\nu\lambda\rho}(\bx-\bmu)\cU_\mu(\bx-\bmu)\right)\eeq
In the limit of zero coupling $\cU=I$ this takes the form
\beq
\frac{1}{3!}\sum_{\bx}\Tr \theta_{\nu\lambda\rho}(\bx)D^{(-)}_\mu 
\kappa_{\mu\nu\rho\lambda}(\bx)\eeq
which now has the correct canonical form to exclude
doubles according to the theorem of Rabin
\cite{Rabin:1981qj}. Notice that the original forward difference has become
a backward difference operator after the change of variables.

The only other term requiring this more careful analysis arises from
the $\cQ$-closed term. The problematic term looks like
\beq
\epsilon_{abcd5}\chi_{5d}(\bx+\bmu_a+\bmu_b+\bmu_c)
\left(\chi_{12}(\bx+\bmu_c)\cUb_c(\bx)-
\cUb_c(\bx+\bmu_a+\bmu_b)\chi_{ab}(\bx)\right)\eeq
Using the result $\sum_{a=1}^5\bmu_a=\bzero$ 
this can be written for zero coupling ($\cU\to I$) as 
\beq
\frac{1}{3!}\theta_{abc}(\bx)D^{(+)}_{\left[c\right.}\chi_{\left.ab\right]}(\bx)\eeq
where the presence of the epsilon symbol ensures the
complete antisymmetrization of the derivative. This final form has the 
form required by Rabin's theorem. The theory is thus manifestly free of doubles.

\section{Conclusions}
In this paper we have shown how to derive the supersymmetric orbifold
lattices of Cohen et al. \cite{Cohen:2003xe} and Kaplan et al.
\cite{Kaplan:2005ta} by geometrical discretization of the continuum
twisted supersymmetric Yang-Mills theory. This connection is not
unexpected -- Unsal showed earlier \cite{Unsal_rel}
that the naive continuum limit of the $\cQ=16$
orbifold theory in four dimensions corresponded to the Marcus twist of
${\cal N}=4$ and more recent work by Damgaard et al.
\cite{Damgaard:2007xi} and Takimi \cite{Takimi:2007nn} have exhibited
the strong
connections between discretizations of the twisted theory and orbifold
theories. Our new work makes the connection complete -- the two approaches
are in fact identical provided one chooses the exact lattice supersymmetry
carefully and uses the geometric discretization proposed in
\cite{Catterall_n=2}. In fact, as was pointed out in \cite{Damgaard:2007eh}
this lattice theory is essentially equivalent to the one proposed
in \cite{D'Adda_2d} {\it provided} that the fermionic shift parameter
employed in that model is chosen to be zero and we restrict 
our
attention solely to the corresponding scalar supercharge.

The case of $\cQ=16$ is particularly interesting. We have shown that the
continuum theory can be recast as the dimensional reduction of a very
simple five dimensional theory. The $\cQ$-exact part of the action is
essentially identical to the two dimensional theory with the primary difference
between the two theories arising because of the appearance of a new
$\cQ$-closed term which was not possible in two dimensions. Nevertheless
discretization proceeds along the same lines, the one subtlety being the
lattice link assignment of the fifth component of the complex gauge field
after dimensional reduction. The key requirement governing
discretization is that successive components
of the \KD field representing the fermions have opposite orientations.
This allows the fermionic action to be gauge invariant without any
additional doubling of degrees of freedom. It seems likely that all
the orbifold actions in various dimensions can be obtained in this manner.

\acknowledgments The author is supported in part by DOE grant
DE-FG02-85ER40237 and would like to acknowledge useful conversations
with Poul Damgaard,  Joel Giedt, David Kaplan, So Matsuura,
Mithat Unsal and Toby Wiseman.

%-----------------------------------------------------------------
%
\bibliographystyle{JHEP}
\bibliography{twist}
%
%-----------------------------------------------------------------

\end{document}